\documentclass[twocolumn]{revtex4-1}

\usepackage{amsmath}
\usepackage{dsfont}
\usepackage{setspace}
\usepackage{graphicx} 
\usepackage{capt-of} 
\usepackage{amstext}
\usepackage[ngerman,english]{babel}
\usepackage{xcolor}
\usepackage{booktabs}
\usepackage{hyperref} 
\hypersetup{colorlinks=false, citebordercolor=green}
\usepackage{soul}

\begin{document}
\title{Doping a lattice-trapped bosonic species with impurities:\\ From ground state properties to correlated tunneling dynamics}

\author{Kevin Keiler $^1$}
\author{Simeon I. Mistakidis $^1$}
\author{Peter Schmelcher $^{1,2}$}
\affiliation{$^1$Center for Optical Quantum Technologies, Department of Physics, University of Hamburg, Luruper Chaussee 149, 22761 Hamburg, Germany}
\affiliation{$^2$The Hamburg Centre for Ultrafast Imaging, University of Hamburg, Luruper Chaussee 149, 22761 Hamburg, Germany}


\begin{abstract}
We investigate the ground state properties and the nonequilibrium dynamics of a lattice trapped bosonic mixture consisting of an impurity species and a finite-sized medium.  For the case of one as well as two impurities we observe that, depending on the lattice depth and the interspecies interaction strength, a transition from a strongly delocalized to a
localized impurity distribution occurs.  In the latter regime the two species phase separate, thereby forming a particle-hole pair. For two impurities we find that below a critical lattice depth they are delocalized among two neighboring outer lattice wells and are two-body correlated. This transition is characterized by a crossover from strong to a suppressed interspecies entanglement for increasing impurity-medium repulsion. 
Turning to the dynamical response of the mixture, upon quenching the interspecies repulsion to smaller values, we reveal that the predominant tunneling process for a single impurity corresponds to that of a particle-hole pair, whose dynamical stability depends strongly on the quench amplitude. During the time-evolution a significant increase of the interspecies entanglement is observed, caused by the build-up of a superposition of states and thus possesses a many-body nature. In the case of two bosonic impurities the particle-hole pair process becomes unstable in the course of the dynamics with the impurities aggregating in adjacent lattice sites while being strongly correlated. 
\end{abstract}

\maketitle

\section{Introduction}
\label{sec:intro}

Ultracold atomic physics offers an excellent testbed for probing the static properties and in particular the nonequilibrium quantum dynamics of multicomponent systems for both fermions and bosons \cite{Inouye,Fukuhara_mixt}. 
It provides an exquisite level of control of several system parameters including, for instance, the intra- and intercomponent scattering lengths via Feshbach resonances \cite{feshbach,Inouye}, the shape of the external trapping potential \cite{boshier,Grimm_trap} as well as the particle number with remarkable experimental achievements especially in one spatial dimension \cite{few,few1}. 

Recently, a major focus has been placed on the study of highly particle imbalanced setups \cite{massignan_review,Kohstall,Koschorreck,Scazza} namely impurities in a many-body environment. 
In this context, the presence of intercomponent interactions results in the dressing of the impurities by the excitations of their medium giving rise, among others, to the concept of quasiparticles \cite{Landau} e.g. polarons \cite{Schmidt_rev,massignan_review,Grusdt_rev}. 
The latter exhibit extraordinary features such as an effective mass \cite{grusdt,Khandekar,Ardila_mass} and induced interactions \cite{keiler1,jie_chen}. 
Owing to the very recent experimental realization of these impurity systems \cite{catani,fukuhura,Kohstall,Koschorreck,Scazza}, an intense theoretical activity has been triggered for the investigation of their stationary properties \cite{volosniev} e.g. unveiling their excitation spectra \cite{Koschorreck,Schmidt_rev,Cetina_intef,Tajima_spec}, induced-interactions \cite{zinner_pol,simosfermi} and self-localization \cite{Sacha_loc}.
However, their corresponding nonequilibrium dynamics still remains largely unexplored. 
This partly stems from the fact that the impurities consist few-body subsystems and thus correlation-induced phenomena are expected to be pronounced especially during the dynamics, which has also been experimentally confirmed \cite{Koepsell_cor}. 
Notable examples here involve, for instance, nonlinear pattern formation \cite{Mistakidis_BF,grusdt}, 
induced-correlations \cite{Mistakidis_indcor,Mistakidis_2impcor,Tajima_col}, relaxation processes \cite{Mistakidis_PPS,Boyanovsky,Lausch,Mistakidis_catastr}, collisional aspects of an impurity with its host \cite{Mistakidis_diss,Mukherjee_pulse,Gamayun,Mathy} as well as tunneling dynamics of impurities in optical lattices \cite{Johnson,Cai,trans_imp,Bruderer,Bruderer_strong_coupl,Bruderer_selftrap,Theel,Keiler_tunel,bf2018}. 
In this latter context transport properties of impurities \cite{Johnson,Cai,trans_imp,Bruderer,Bruderer_strong_coupl}, self-trapping phenomena \cite{Bruderer_selftrap,Yin_selftrap} and Bloch oscillations \cite{Grusdt_bloch} have been evinced.  

However, the majority of these lattice trapped impurity investigations have been mainly focusing on the case that only the impurities experience the lattice potential and the host resides in a homogeneous environment. 
Moreover, they have been predominantly restricted to the single impurity case \cite{duncan_imp} and operated within the lowest-band approximation \cite{torma_imp1, torma_imp2}. 
Thus, the situation where both the impurities and their medium are trapped in the same lattice potential remains an open question. 
In such a setting the impurities act as defects possessing a particle character and it would be intriguing to study the different phases that arise in the ground state of this composite system for variable impurity-medium interactions and unveil their underlying correlation properties. 
Recall that lattice trapped particle-balanced bosonic mixtures exhibit quantum phases \cite{Guglielmino,Buonsante,Gadway,Thalhammer,Kato} being absent in their single-component counterpart. 
This is, partly, caused by the non-negligible presence of interspecies correlations \cite{Wang_ent_lat}. 
For instance, modifications of the Mott-insulator (MI) to the superfluid (SF) phase transition \cite{Guglielmino,Buonsante,Gadway,Thalhammer,imm_misc_temp} have been reported due to the existence of a second component leading to the so-called paired and counterflow superfluid states \cite{Hu,Hu_noise}, quantum emulsion states \cite{Roscilde,Buonsante} as well as losses of the intracomponent coherence \cite{Catani_coh}. 
In this sense, it is natural to investigate the existence and interplay of the different phases in the particle imbalanced scenario with respect to the interspecies interaction strength.   
As a prototypical example, henceforth we consider one or two bosonic impurities immersed in a majority species of bosons with both components being lattice trapped in one-dimension. 

Having established the ground state properties of this setup another fruitful prospect is to inspect its corresponding nonequilibrium dynamics by quenching the system between the different emergent phases. Here the analysis and consequent control of the tunneling dynamics of the impurities is of particular importance since it might give rise to a variety of complex transport phenomena, self-trapping events and formation of (repulsively) bound pairs \cite{Theel,Keiler_tunel,bf2018,Yin_selftrap,zollner1,zollner2}. 
Furthermore, the identification of the correlated many-body nature of the different tunneling processes will allow us to infer their microscopic origin which is certainly of interest. 
To track the static properties and the quench dynamics of the particle imbalanced Bose-Bose mixture we utilize the variational Multi-Layer Multi-Configuration Time-Dependent Hartree method for atomic mixtures (ML-MCTDHX) \cite{mlb1,mlb2,mlx} which enables us to capture all the relevant interparticle correlations of this multicomponent setup. 

Regarding the ground state of a single and two bosonic impurities immersed in a majority bosonic species we find a transition from a SF to a MI phase of the composite system (doped insulator) for a specific lattice depth and increasing interspecies repulsion. This is in sharp contrast to the case of a homogeneous bath where a MI state cannot be naturally achieved. 
This transition takes place for weaker impurity-medium interactions for deeper lattices, a result which is more pronounced in the two impurity case. 
Within the SF phase the impurity and the majority species show a delocalized behavior with the interspecies entanglement being enhanced and the medium being characterized by strong two-body correlations.  
However entering the MI state of the mixture, the species phase separate forming a particle hole-pair and their entanglement is suppressed \cite{Mistakidis_phase_sep,Theel}. The formation of the particle hole-pair is exclusively caused by the impurity acting as a defect for the bath, an effect being absent in particle-balanced mixtures due to their similar intrinsic composition. 
In this latter case the many-body state of the system exhibits a two-fold degeneracy \cite{keiler2}.  
Moreover, for two impurities we observe that below a critical lattice depth the impurities are delocalized among two neighboring outer wells of the lattice and are two-body correlated. 

Turning to the dynamical response of the mixture, upon quenching the interspecies repulsion from a MI to a SF phase, we reveal that the predominant tunneling process for a single impurity corresponds to that of a particle-hole pair \cite{Endres_par_hole}, whose dynamical stability depends strongly on the quench amplitude.  
More specifically, the initially localized impurity becomes spatially delocalized in the course of the evolution while it gradually tunnels from one side of the lattice to the other. 
On the other hand, the majority species particles tend to avoid the impurity in the course of the tunneling.  
During the time-evolution a significant increase of the interspecies entanglement is observed, which is due to the build-up of superposition of states and thus possesses a many-body nature. 
Additionally, strong correlations occur between the particles of the majority species. 
In the case of two bosonic impurities the particle-hole pair process becomes unstable during the evolution with the impurities aggregating in adjacent lattice sites while being strongly correlated. 

This work is structured as follows. 
In sec. \ref{sec:setup_methodology} we introduce our setup and discuss the variational many-body approach. 
Sec. \ref{sec:Ground_State_Properties} presents the ground state properties in a finite lattice for a single and two impurities immersed in a strongly interacting majority bosonic species with filling smaller than unity. 
The nonequilibrium dynamics of the impurities by quenching the interspecies interaction strength from the doped insulator to the SF phase is analyzed in sec. \ref{sec:Tunneling_Dynamics}. 
We summarize our results and provide an outlook in sec. \ref{sec:Conclusion}. 
Appendix \ref{sec:Appendix1} elaborates on the lattice trapped ground state phase diagram of an impurity in a unit filling majority species.

\section{Setup and variational many-body approach}
\label{sec:setup_methodology}

\subsection{Treatment of Many-Body Correlations and Dynamics}
Our computation approach is the \textit{ab-initio} Multi-Layer Multi-Configuration Time-Dependent Hartree method for bosonic (fermionic) Mixtures (ML-MCTDHX) \cite{mlb1,mlb2,mlx}, which accounts for all the relevant correlations of the atomic mixture \cite{Mistakidis_phase_sep,ff2018,ff2019,bf2018,lode1,lode2}.
As a first step, the total many-body wave function $|\Psi_{\textrm{MB}}(t)\rangle$ is expanded in $M$ species functions $|\Psi^{\sigma}(t)\rangle$ of species $\sigma$ and written as a Schmidt decomposition \cite{schmidt_dec} 
\begin{equation}
|\Psi_{\textrm{MB}}(t)\rangle = \sum_{i=1}^{M} \sqrt{\lambda_{i}(t)} |\Psi_{i}^A(t)\rangle\otimes |\Psi_{i}^B(t)\rangle.
\label{eq:schmidt}
\end{equation}
Here, the Schmidt coefficients $\sqrt{\lambda_{i}}$, in decreasing order, provide information about the degree of population of the $i-$th species function and thereby determine the degree of entanglement between the impurities and the majority species. In case that $\lambda_1=1$ the species $A$ and $B$ are not entangled and the system can be described with a species mean-field ansatz corresponding to a single product state ($M=1$).

Furthermore, the species wave functions $|\Psi^{\sigma}(t)\rangle$ describing an ensemble of $N_\sigma$ bosons are expanded in a set of permanents
\begin{equation}
|\Psi_{i}^\sigma(t)\rangle = \sum_{\vec{n}^\sigma|N_\sigma} C_{\sigma\vec{n}}(t)
 |\vec{n}^\sigma;t\rangle,
\label{eq:ml_ns}
\end{equation}
where the vector $\vec{n}^\sigma=(n^{\sigma}_1,n^{\sigma}_2,...)$ denotes the occupations of the time-dependent single-particle functions of the species $\sigma$. The notation $\vec{n}^\sigma|N_\sigma$ indicates that for each $|\vec{n}^\sigma;t\rangle$ we require the condition $\sum_{i}n^{\sigma}_i=N_\sigma$.
The time propagation of the many-body wave function is achieved by employing the Dirac-Frenkel variation principle $ \langle\delta\Psi_\textrm{MB}| (\textrm{i}\partial_t - \mathcal{H} )|\Psi_\textrm{MB}\rangle $ \cite{var1,var2,var3} with the variation $\delta\Psi_\textrm{MB}$.
ML-MCTDHX provides access to the complete many-body wave function which allows us consequently to derive all relevant characteristics of the underlying system. In particular, this means that we are able to characterize the system by projecting onto number states with respect to an appropriate single-particle basis \cite{ns_analysis1,ns_analysis2}. Besides investigating the quantum dynamics it allows us to determine the ground (or excited) states by using either imaginary time propagation or improved relaxation \cite{meyer_improved}, thereby being able to uncover also possible degeneracies of the many-body states. We remark that in commonly used approaches for solving the time-dependent Schr{\"o}dinger equation, one typically constructs the wave function as a superposition of time-independent Fock states with time-dependent coefficients. Instead, it is important to note that the ML-MCTDHX approach considers a co-moving time-dependent basis on different layers, meaning that in addition to time-dependent coefficients the single particle functions spanning the number states are also time-dependent. This leads to a significantly smaller number of basis states and configurations that are needed to obtain an accurate description of the system under consideration and thus reduces the computation time \cite{pruning}. 

The degree of truncation of the underlying Hilbert space is given by the orbital configuration $C=(M,d_A,d_B)$. Here, $M$ refers to the number of species functions in the Schmidt decomposition (cf. equation \ref{eq:schmidt}), while $d_\sigma$ with $\sigma \in\{A,B\}$ denote the number of single-particle functions spanning the time-dependent number states $ |\vec{n}^\sigma; t\rangle$ (cf. equation \ref{eq:ml_ns}). The orbital configuration $C=(7,7,7)$ has been employed for all many-body calculations presented in the main text, yielding a converged behavior of our observables.

\subsection{Lattice Trapped Bosonic Mixture}
\begin{figure}[t]
	\includegraphics[width=0.7\columnwidth]{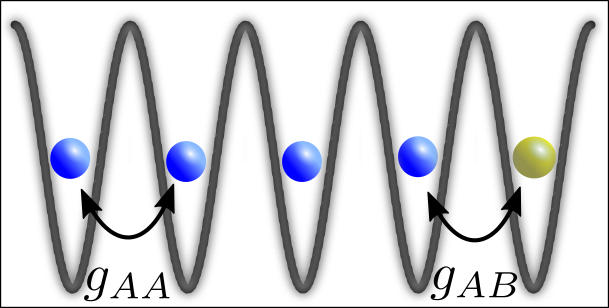}
	\caption{(a) Sketch of the two-component lattice trapped bosonic mixture. The majority species atoms (blue spheres) interact repulsively via an intraspecies contact interaction of strength $g_{AA}$ and an interspecies repulsion of strength $g_{AB}$ with the bosonic impurities.}
	\label{fig:initial_state}
\end{figure}
Our system consists of a mixture of two bosonic species which are trapped in a one-dimensional lattice with hard wall boundary conditions at its endpoints [see Figure \ref{fig:initial_state}]. The impurity species with $N_B=1,2$ particles is denoted as species B and the majority species containing $N_A=4$ (main text) or $N_A=5$ (appendix) particles is referred to as species A. This setup lies within reach of current experimental techniques \cite{spec_sel_lat,box_pot}. Furthermore, we introduce a coupling Hamiltonian $\hat{H}_{AB}$ between the two species. Both subsystems are confined along the longitudinal spatial direction, accounting for the one-dimensional character of our setup. Excitations in the corresponding transversal direction are energetically suppressed in the scenario under investigation and can therefore be neglected in our setup. This results in a Hamiltonian of the form $\hat{H}=\hat{H}_A+\hat{H}_B+\hat{H}_{AB}$.
The Hamiltonian of the species $\sigma$, with $\sigma \in \{A,B\}$, reads
\begin{equation}
\begin{split}
\hat{H}_\sigma&=\int_{-L/2}^{L/2} \text{dx} \, \hat{\Psi}_{\sigma}^{\dagger}(\text{x}) \Big [ -\frac{\hbar^{2}}{2 m_\sigma} \frac{\text{d}^{2}}{\text{dx}^{2}}+ V_0 \sin^{2}\Big(\frac{\pi k \text{x}}{L}\Big) \\
&+ g_{\sigma\sigma} \; \hat{\Psi}_{\sigma}^{\dagger}(\text{x}) \hat{\Psi}_{\sigma}(\text{x}) \Big ] \hat{\Psi}_{\sigma}(\text{x}),
\end{split}
\end{equation}
where $\hat{\Psi}_{\sigma}^{\dagger}$ is the field operator of species $\sigma$, $m_\sigma$ their mass and $V_0$ the lattice depth. Also, $g_{\sigma\sigma}$ refers to the intraspecies interaction strength of the two-body contact interaction among the $\sigma$ atoms, $k$ is the number of lattice wells and $L$ is the length of the system, while $x\in[-L/2,L/2]$.
Moreover, we assume equal masses for the species $m_A=m_B$. Experimentally this can be achieved by preparing e.g. $^{87}{Rb}$ atoms in two different hyperfine states \cite{mbl_hyperfine}. The interaction between the species A and B is given by
\begin{equation}
\hat{H}_{AB}= g_{AB} \int_{-L/2}^{L/2} \text{dx} \; \hat{\Psi}_{A}^{\dagger}(\text{x}) \hat{\Psi}_{A}(\text{x}) \hat{\Psi}_{B}^{\dagger}(\text{x}) \hat{\Psi}_{B}(\text{x}),
\end{equation}
where $g_{AB}$ is the effective one-dimensional interspecies interaction strength. The interaction strengths $g_{\alpha}$ ($\alpha\in\{A,B,AB\}$) can be expressed in terms of the three dimensional s-wave scattering lengths $a^{3D}_{\alpha}$. By assuming the above-mentioned strong transversal confinement with the same trapping frequencies $\omega^{\sigma}_{\perp}=\omega_{\perp}$ for both species $\sigma \in \{A,B\}$ it is possible to integrate out frozen degrees of freedom, leading to a quasi one-dimensional model with $g_{\alpha}=2\hbar\omega_{\perp}a^{3D}_{\alpha}$.\par
Throughout this work we consider a $k=5$ well lattice, while the interaction among the majority atoms is set to a value where the particles distribute in a Mott-like state for large lattice depths, namely $g_{AA}/E_R \lambda=0.04$. Here, $E_R=(2\pi\hbar)^{2}/2m_A \lambda^{2}$ is the recoil energy and $ \lambda=2L/k$ the optical lattice wavelength.
\section{Ground State Properties}
\label{sec:Ground_State_Properties}
\begin{figure*}[t]
	\includegraphics[scale=0.35]{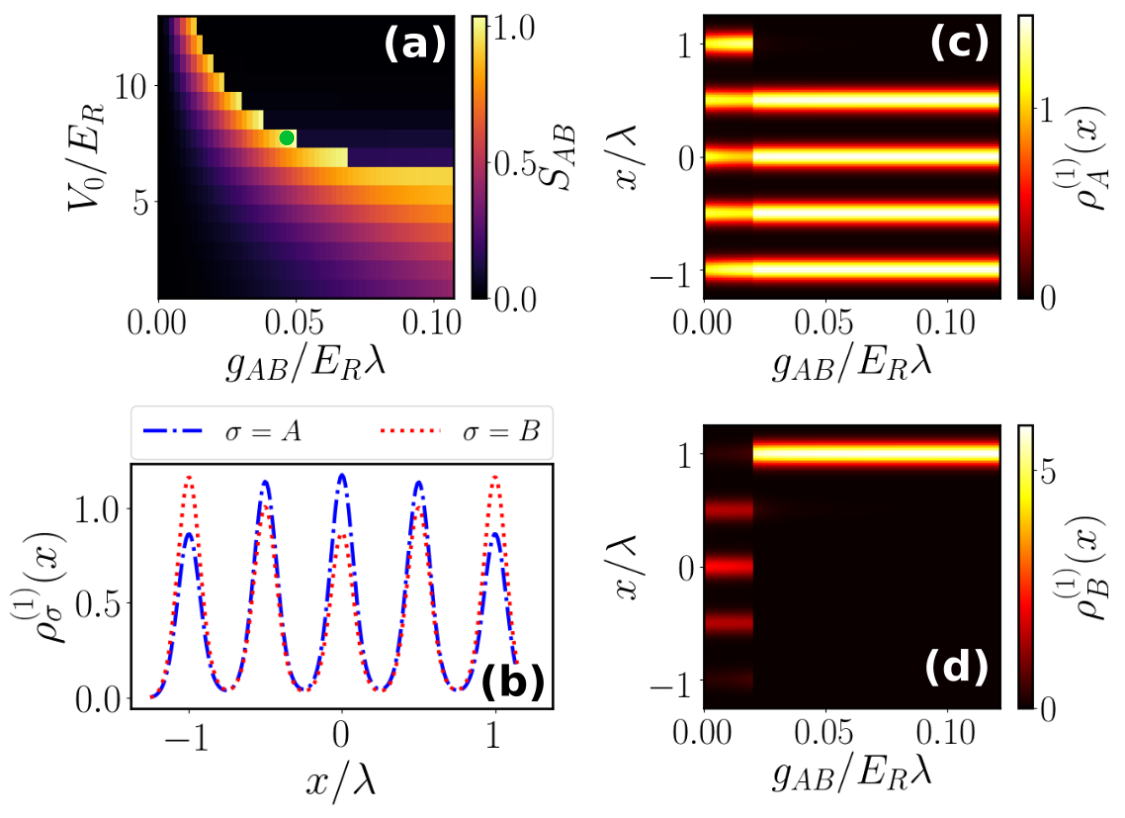}
	\caption{(a) The  von Neumann entropy $S_{AB}$ as a function of the interspecies interaction strength $g_{AB}$ and the lattice depth $V_0$. (b) One-body density $\rho^{(1)}_\sigma(x)$ of the $\sigma$-species for $V_0/E_R=7.3$ and $g_{AB}/E_R\lambda=0.047$. This density distribution corresponds to the case where the von Neumann entropy is very large [see green dot in panel (a)]. One-body density $\rho^{(1)}_\sigma(x)$ of (c) the species A and (d) of the species B in dependence of the interspecies interaction strength $g_{AB}$ for a fixed lattice depth of $V_0/E_R=10.5$. The particle number of the respective species is $N_A=4$ and $N_B=1$.}
	\label{fig:gs_densities_single_imp}
\end{figure*}
Let us analyze the ground state properties of the Bose-Bose mixture with respect to the lattice depth $V_0$ and the interspecies coupling strength $g_{AB}$ for $N_B=1$ and $N_B=2$ impurities. We calculate the many-body ground state of the Bose-Bose mixture using ML-MCTDHX, which enables us to obtain the resulting full many-body wave function. In order to be able to interpret the wave function, we shall analyze reduced quantities such as the von Neumann entropy and the one- and two-body densities of each species on the basis of the numerically obtained many-body wave function. As a result, we are able to gain an in-depth insight into the spatial distribution of the two species in the lattice potential and the accompanying intra- and inter-species correlations.
\subsection{Single Impurity}
\label{sec:Single_Impurity}
In the following, we explore the ground state of the system containing a single impurity, i.e. $N_B=1$, for varying $V_0$ and $g_{AB}$, while keeping fixed the intraspecies interaction strength to $g_{AA}/E_R \lambda=0.04$. As a first step, we analyze the spatial distribution of the two species in terms of the one-body density of the ground state $|\Psi_{\text{MB}}\rangle$ of the species $\sigma$, which is defined as
\begin{equation}
\rho^{(1)}_\sigma(x)=\langle\Psi_\text{MB}| \hat{\Psi}_{\sigma}^{\dagger}(x)\hat{\Psi}_{\sigma}(x)|\Psi_\text{MB} \rangle.
\label{eq:obd}
\end{equation}
Additionally, in order to deepen our understanding of the ground state of the binary mixture, we investigate the degree of correlations and, in particular the entanglement between the impurity species and the majority species.
For this purpose, we introduce the von Neumann entropy
\begin{equation}
S_{AB}=-\sum_{i} \lambda_i \ln(\lambda_i)
\label{eq:vonNeumann}
\end{equation}
as a measure for the entanglement between the subsystems A and B, where $\lambda_i$ are the Schmidt coefficients defined in equation \ref{eq:schmidt}. Recall that in the case of a single contributing product state in equation \ref{eq:schmidt}, the subsystems are disentangled and the von Neumann entropy is $S_{AB}=0$, whereas any deviation from this value indicates entanglement between the A and the B species.

Figure \ref{fig:gs_densities_single_imp} (a) shows the von Neumann entropy of the ground state as a function of the interspecies interaction strength $g_{AB}$ and the lattice depth $V_0$. For small $g_{AB}$ we observe that $S_{AB}\approx0$, which indicates that our system is well described by a single product state. Increasing $g_{AB}$ leads to a growth of the von Neumann entropy. For sufficiently large lattice depths $S_{AB}$ is maximized for a specific value of $g_{AB}$ (cf.  $V_0 /E_R = 7.3$ and $g_{AB} /E_R\lambda= 0.047$), while any further increase of the latter leads to a sudden reduction of the entropy becoming subsequently close to zero which again corresponds to a single product state representation of the many-body wave function.

In order to understand the relationship between the particle distribution of each species and the von Neumann entropy it is useful to investigate the one-body density of the $\sigma$-species as a function of the interspecies interaction strength [see Figure \ref{fig:gs_densities_single_imp} (c), (d)]. As it can be seen, up to a specific value $g_{AB}$ the majority species A is distributed over the whole lattice geometry, with a slight decrease (increase) of the density in the central (outer) well(s) for a larger $g_{AB}$. From this critical value onward the majority species forms a hole in one of the outer wells and it is now only distributed over the four remaining wells. Similarly, we observe for the impurity that up to this critical value of $g_{AB}$ it is distributed over the central three sites, showing an increasing density in the outer wells for larger $g_{AB}$. However, as soon as a hole is formed in the majority species the impurity localizes in a single outer well which is unoccupied by the majority species. The latter is accompanied by the formation of a two-fold degeneracy in the ground state. In this sense, the ground state is given by the density distribution for large $g_{AB}$ as depicted in figure \ref{fig:gs_densities_single_imp} (c), (d) and its parity-symmetric (with respect to $x=0$) counterpart. Consequently, the densities shown for large $g_{AB}$ correspond to only one of the two energetically degenerate ground states.
Focusing on the above-described critical value of $g_{AB}$, e.g. for $V_0 /E_R$ = 10.5 we observe a minor population of the impurity in the outer wells. However, this spatial species distribution in $\rho^{(1)}_B(x)$ is more pronounced for smaller values of $V_{0}$. This means that the corresponding one-body density is increased in the outer wells. Indeed, for $V_0 /E_R = 7.3$ and $g_{AB} /E_R \lambda= 0.047$ [as compared to $V_0 /E_R =10.5$ and $g_{AB} /E_R\lambda= 0.02$ in Figure \ref{fig:gs_densities_single_imp} (d)] the impurity is largely distributed over the lattice geometry and exhibits an increased density in the outer wells [see Figure \ref{fig:gs_densities_single_imp} (b)]. Correspondingly, the density of the majority species is smaller in the outer wells compared to the central ones. This large overlap between the two species on the level of the one-body densities is responsible for the maximized von Neumann entropy at the critical value of $g_{AB}$. In turn, this explains the sudden decrease of $S_{AB}$ which is associated with a phase-separation of the two species \cite{phase_sep1,ofir,phase_sep2,phase_sep3} and the formation of a doped insulator for the composite system. 
Concluding, an increase of the von Neumann entropy is associated with a strong delocalization of the impurity, thereby leading to a large overlap of the two species. From a critical value of $g_{AB}$ onward we observe a phase-separation of the two species which is accompanied by a decrease of the von Neumann entropy to $S_{AB}\approx0$. This can also be viewed as the formation of a particle-hole pair \cite{Endres_par_hole}, where the majority species forms the hole in one of the outer wells. 
\begin{figure*}[t]
	\includegraphics[scale=0.35]{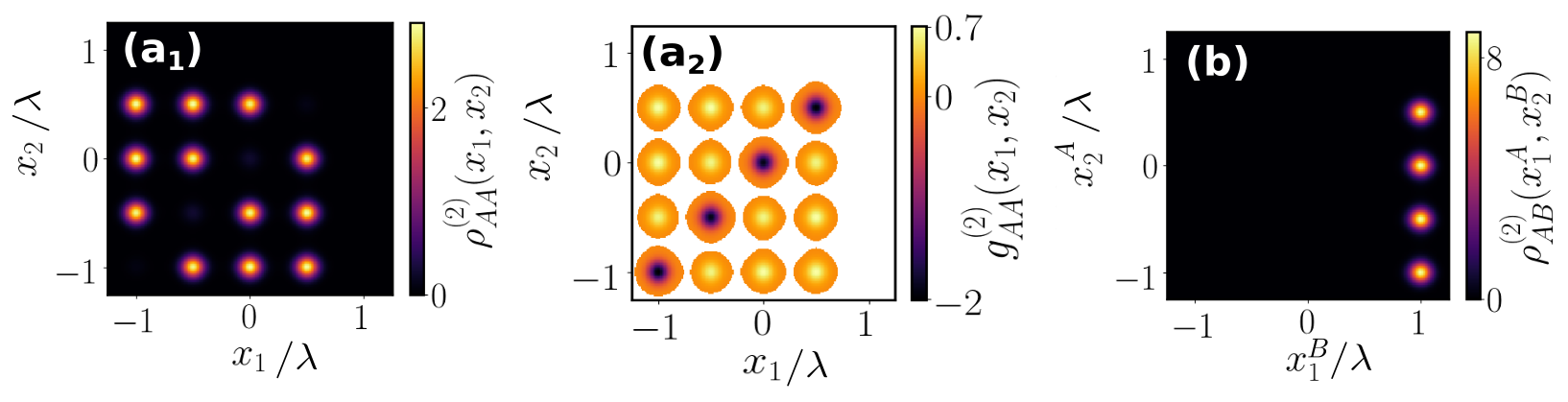}
	\caption{The two-body density for ($a_1$) two particles of the majority species $\rho^{(2)}_{AA}$ and (b) for one particle of the majority species and one being an impurity $\rho^{(2)}_{AB}$. ($a_2$) The noise-correlation for two particles of the majority species $\rho^{(2)}_{AA}$. We consider $V_0/E_R=10.5$ and $g_{AB}/E_R\lambda=0.12$. The particle number of the respective species is $N_A=4$ and $N_B=1$.}
	\label{fig:g2_rho2_single_imp}
\end{figure*}

To obtain a deeper understanding of the particle distribution, as a next step, we inspect the two-body reduced density which reads
\begin{align}
\label{eq:two_body}
&\rho^{(2)}_{\sigma\sigma^{\prime}}(x^\sigma_1,x^{\sigma^{\prime}}_2)=\\ \nonumber
&\langle\Psi_\text{MB}|\hat{\Psi}_{\sigma}^{\dagger}(x^\sigma_1)\hat{\Psi}_{\sigma^{\prime}}^{\dagger}(x^{\sigma^{\prime}}_2)\hat{\Psi}_{\sigma}(x^\sigma_1)\hat{\Psi}_{\sigma^{\prime}}(x^{\sigma^{\prime}}_2)|\Psi_\text{MB} \rangle,
\end{align}
where $\sigma,\sigma^{\prime} \in \{A,B\}$. This measure corresponds to the probability of finding a particle of species $\sigma$ at the position $x^\sigma_1$ and another particle of species $\sigma^{\prime}$ at the position $x^{\sigma^{\prime}}_2$ \footnote{Please note, that in the case two particles are of the same species, we omit the superscript of the species label at the position symbol.}. Figure \ref{fig:g2_rho2_single_imp} ($a_1$) illustrates this quantity for two particles of the majority species and Figure \ref{fig:g2_rho2_single_imp} (b) for one particle of the majority species and one impurity for $V_0/E_R=10.5$ and $g_{AB}/E_R\lambda=0.12$. This parameter set corresponds to the case where the two species form a particle hole-pair and thereby phase-separate. This fact can also be observed in the behavior of the interspecies two-body density $\rho^{(2)}_{AB}$ [see Figure \ref{fig:g2_rho2_single_imp} (b)]. Indeed, the probability of finding a particle of species A and a particle of species B at the same position is approximately zero, i.e. $\rho^{(2)}_{AB}(x,x)\approx0$. Instead, measuring the impurity in one of the outermost wells, we may find the majority species localized in any other well with approximately the same probability. This indicates that the majority species is equally distributed over four out of the five wells. In this context, the question arises how the majority species particles distribute among each other. As shown, in Figure \ref{fig:g2_rho2_single_imp} ($a_1$) we find that the probability of detecting two particles of species A at the same position is approximately zero, i.e. $\rho^{(2)}_{AA}(x,x)\approx0$. This means that two particles of the majority species tend to avoid each other and do not occupy the same well. From this we can conclude, that they form a Mott insulator-like state on the four populated wells. 
In this sense, the complete wave function of the system in this regime can be well described as follows
\begin{equation}
|\Psi_{\textrm{MB}}\rangle \approx |1,1,1,1,0\rangle_A\otimes|0,0,0,0,1\rangle_B.
\label{eq:single_imp_wfn}
\end{equation}
Here, the number state $|\vec{n}^{\sigma}\rangle=|n^\sigma_1,n^\sigma_2,n^\sigma_3,n^\sigma_4,n^\sigma_5\rangle_\sigma$ is constructed with a generalized Wannier basis of the lowest band \cite{kivelson1,kivelson2}. This means that e.g. $n^\sigma_1$ describes the number of $\sigma$ atoms in the left localized Wannier state of the lowest band. This state essentially describes the doped insulator configuration of the composite system. 

In order to unravel the role of correlations for the ground state distribution in the strong interaction regime, we calculate the noise correlation \cite{Hu_noise} between particles of species $\sigma$ and $\sigma^{\prime}$ which is defined as
\begin{equation}
g^{(2)}_{\sigma\sigma^{\prime}}(x^\sigma_1,x^{\sigma^{\prime}}_2)=\rho^{(2)}_{\sigma\sigma^{\prime}}(x^\sigma_1,x^{\sigma^{\prime}}_2)-\rho^{(1)}_\sigma(x^\sigma_1)\rho^{(1)}_\sigma(x^{\sigma^{\prime}}_2).
\label{eq:noise_correlation}
\end{equation}
The noise correlation is a measure for the deviation of the conditional probability of finding two particles at specific positions from the unconditional one given by the product of two single-particle events. In this sense, it gives insight into whether two particles can be viewed as independent from each other or not and therefore suggest the occurrence of beyond single-particle processes in the system. In the former case $g^{(2)}_{\sigma\sigma^{\prime}}=0$ and in the latter case $g^{(2)}_{\sigma\sigma^{\prime}} \neq0$. 
In Figure \ref{fig:g2_rho2_single_imp} ($a_2$) we show  $g^{(2)}_{AA}=0$ corresponding to the previously discussed two-body density for two particles of the A species for $V_0/E_R=10.5$ and $g_{AB}/E_R\lambda=0.12$. Let us remark that in case of $g^{(2)}_{\sigma\sigma^{\prime}}<10^{-2}$, two-body processes are negligible and therefore we set the color of the colormap to white. Moreover, we do not show the interspecies noise correlation, since it does not exhibit any significant structures due to the ground state being well described by the wave function of equation \ref{eq:single_imp_wfn}. The measurement of one particle of species A does not depend on the previous measurement of the particle in species B and vice versa. However, as one might expect the noise correlation between two particles of species A shows a more involved structure. In particular the conditional probability of finding two particles of the majority species at the same position deviates strongly from the unconditional one. This is a clear signature of the Mott insulator-like state formed by the A species. Furthermore, we find a difference between the two probabilities in the off-diagonal elements of $g^{(2)}_{AA}$, which again emphasizes the correlated nature of this state. 

In summary, we have found that our system containing a single impurity interacting repulsively with a majority species undergoes a transition regarding the ground state of the system in dependence of the interspecies interaction strength and the lattice depth. This transition manifests itself in an increase of interspecies entanglement with increasing $g_{AB}$, which is accompanied by a delocalization of the impurity followed by a sudden decrease of the entanglement. The latter is due to the phase-separation (or particle-hole formation) between the species constituting a doped insulator, which takes place for sufficiently large $g_{AB}$ and exhibits an energetic degeneracy of the ground state.

\subsection{Two Bosonic Impurities}
\label{sec:Two_Impurities}
In the following, we investigate the ground state of the above-discussed system, but with an additional impurity, i.e. $N_B=2$. In order to focus on the effect of the interspecies interaction we set the intraspecies interaction strength among the impurity particles to zero, i.e. $g_{BB}=0$. Analogously to the analysis above we first examine the interspecies entanglement and the corresponding one-body densities of each species.
\begin{figure*}[t]
	\includegraphics[scale=0.35]{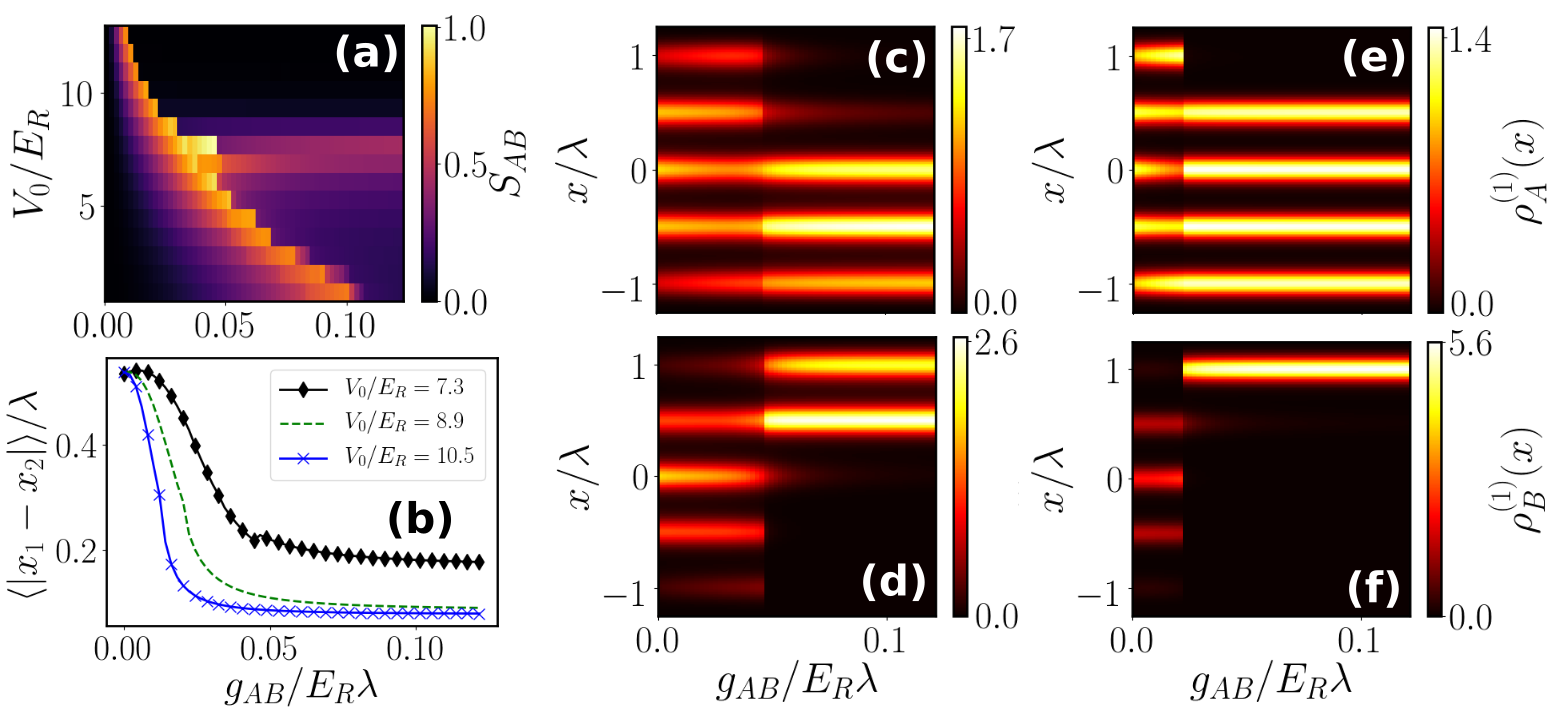}
	\caption{(a) The  von Neumann entropy $S_{AB}$ as a function of the interspecies interaction strength $g_{AB}$ and the lattice depth $V_0$. (b) Distance $\langle|x_1-x_2|\rangle$ of the two impurities as a function of $g_{AB}$ for various $V_0$. One-body density $\rho^{(1)}_A(x)$ of the species A for a fixed lattice depth of (c) $V_0/E_R=5.7$ and (e)  $V_0/E_R=8.9$. One-body density $\rho^{(1)}_B(x)$ of the impurities in dependence of the interspecies interaction strength for a fixed lattice depth of (d) $V_0/E_R=5.7$ and (f)  $V_0/E_R=8.9$. The particle number of the respective species is $N_A=4$ and $N_B=2$.}
	\label{fig:gs_densities_two_imp}
\end{figure*}
Evidently, in Figure \ref{fig:gs_densities_two_imp} (a) we observe an alteration of the crossover diagram, given by the von Neumann entropy as a function of $g_{AB}$ and $V_0$, compared to the case of a single impurity [see also Figure \ref{fig:gs_densities_single_imp} (a)]. More specifically, for large lattice depths an increase of entanglement between the species takes place up to a critical value of the interspecies interaction strength followed by a sudden reduction of $S_{AB}$ to zero for a further increase of $g_{AB}$.
\begin{figure*}[t]
	\includegraphics[width=\textwidth]{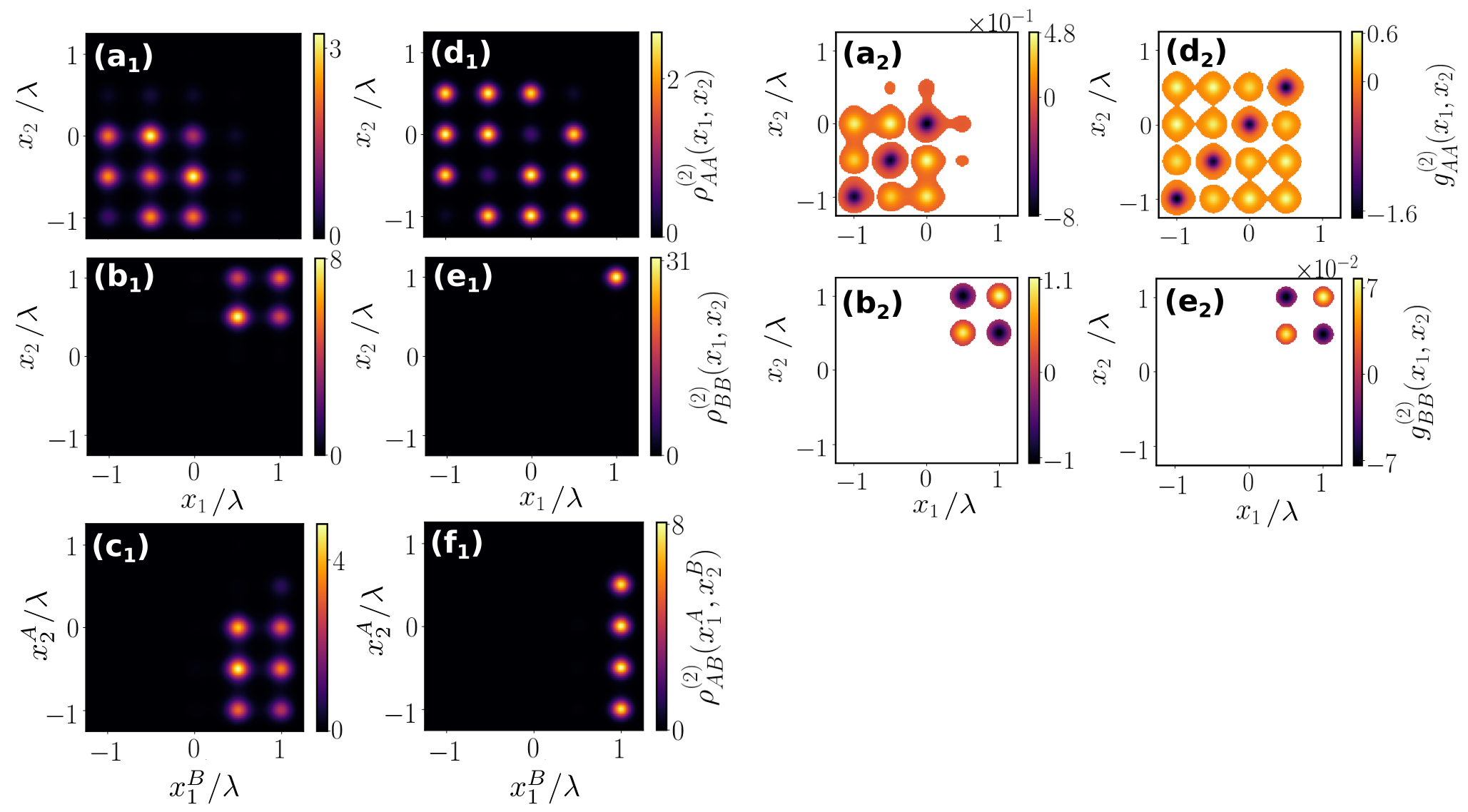}
	\caption{The two-body density for  two particles ($a_1$), ($d_1$) of the majority species $\rho^{(2)}_{AA}$, for ($b_1$), ($e_1$) of the impurity species $\rho^{(2)}_{BB}$ and ($c_1$), ($f_1$) for one particle of the majority species and one impurity $\rho^{(2)}_{AB}$.
	The noise-correlation for two particles ($a_2$), ($d_2$) of the majority species $\rho^{(2)}_{AA}$ and ($b_2$), ($e_2$) of the impurity species $\rho^{(2)}_{BB}$. The column ($a_i$) - ($c_i$) corresponds to a lattice depth of $V_0/E_R=5.7$ and $g_{AB}/E_R\lambda=0.12$, while for ($d_i$) - ($f_i$)  $V_0/E_R=8.9$ and $g_{AB}/E_R\lambda=0.12$, with $i\in\{1,2\}$. The particle number of the respective species is $N_A=4$ and $N_B=2$.}
	\label{fig:g2_rho2_two_imp}
\end{figure*}
This increase of entanglement is again related to the delocalization of the impurities in the lattice, thereby enhancing the overlap between the species on the one-body density level [see Figure \ref{fig:gs_densities_two_imp} (f)]. The sudden reduction of $S_{AB}$ in turn is a result of the phase-separation of the two species [see Figure \ref{fig:gs_densities_two_imp} (e),(f)], similar to the case of a single impurity.
However, for smaller values of the lattice depth (e.g. $V_0/E_R=5.7$) we observe a slightly different behavior of the entanglement and the related one-body densities, as compared to the case of a single impurity. We still find a critical $g_{AB}$ which is associated with the delocalization of the impurity species, while for larger values of $g_{AB}$ the entanglement does not drop to zero. Instead, $S_{AB}$ saturates towards a finite value, which means that the two species remain entangled and the ground state cannot be described by a single product state. The reason for this can be traced back to the distribution of the one-body densities of the two species. For large $g_{AB}$ the impurity species distributes over two of the outer wells with an increased density in the well which is closer to the center [see Figure \ref{fig:gs_densities_two_imp} (d)]. The majority species in turn exhibits a residual density in this very well [see Figure \ref{fig:gs_densities_two_imp} (c)], which leads to a finite overlap of the two species and thereby to a finite entanglement. Let us again remark here that the above-discussed two ground states for $V_0/E_R=5.7$ and $V_0/E_R=8.9$ and large $g_{AB}$ [cf. Figure \ref{fig:gs_densities_two_imp} (c)-(f)] are two-fold degenerate with a parity-symmetric counterpart (with respect to $x=0$).

In the next step, we want to understand how the two impurities are distributed. As a first measure, we investigate the relative distance \cite{Mistakidis_BF,Mistakidis_2impcor} of the two impurities
\begin{equation}
\langle|x_1-x_2|\rangle=\int_{-L/2}^{L/2}\int_{-L/2}^{L/2}dx_1dx_2|x_1-x_2|\rho_{BB}^{(2)}(x_1,x_2),
\end{equation}
with  $\rho_{BB}^{(2)}(x_1,x_2)$ being the previously introduced two-body density of two impurities.
Figure \ref{fig:gs_densities_two_imp} (b) shows the impurity distance as  a function of the interspecies interaction strength for various lattice depths. We observe that for increasing $g_{AB}$ and fixed $V_0$ the distance of the impurities decreases and saturates to a specific value for even larger $g_{AB}$. Also, for smaller $V_0$ we identify a larger impurity distance. Apparently, in spite of the delocalization of the impurity species in the one-body density up to a critical value, the distance $\langle|x_1-x_2|\rangle$ decreases, which suggests that the nature of the delocalization is not completely intuitive and needs to be inspected in detail (see below). The saturation of the impurity distance towards a finite value for large $g_{AB}$ can be associated with the localization of the impurities in a single outer well for very deep lattices or in two adjacent sites for smaller $V_0$. The fact that the distance for the latter case should be larger as compared to the former one, becomes evident in the clear separation of the corresponding lines for large $g_{AB}$ [cf. black diamonds and green dashed line in Figure \ref{fig:gs_densities_two_imp} (b)]. In general, note that the decay of the relative impurity distance is indicative of an induced attractive interaction between the impurities \cite{jie_chen,keiler1,zinner_ind,keiler2}. However, this quantity does not allow for a detailed insight into the actual spatial distribution of the individual impurities. This can be gained by investigating the spatially resolved two-body density of the impurities (see equation \ref{eq:two_body}).

Let us in the following focus on the two cases of large $g_{AB}$, where the impurity species either localizes on a single site [cf. Figure \ref{fig:gs_densities_two_imp} (f)] or on two adjacent sites [cf. Figure \ref{fig:gs_densities_two_imp} (d)], a result that depends on the lattice depth.
For the case where the impurities localize on a single site we find that the majority species exhibits mostly a Mott insulator like structure [cf. Figure \ref{fig:g2_rho2_two_imp} ($d_1$)] as in the single impurity scenario. This implies that the particles of the majority species mostly tend to avoid each other and each one occupies a single distinct lattice site. Moreover, in this context the probability $\rho_{BB}^{(2)}$ of finding two impurities at specific positions $x_1$ and $x_2$ accumulates at a single outer site of the lattice geometry  [see Figure \ref{fig:g2_rho2_two_imp} ($e_1$)]. This means that the two impurities tend to occupy the same site. The previously observed phase-separation of the two species is then also reflected in the corresponding two-body density $\rho_{AB}^{(2)}$  [cf. Figure \ref{fig:g2_rho2_two_imp} ($f_1$)]. 
In general, we can conclude that two impurities behave similarly to a single impurity for large lattice depths and large interspecies interaction strengths.
However, the situation is not that intuitive when the impurity species occupies two adjacent sites for smaller lattice depths. Here, the majority species does not exhibit a Mott insulator-like structure, where the particles avoid each other. Instead, two particles of the majority species can be mostly found either at the same site or on two different ones, while the occupied sites are dominantly those which are not populated by the impurity species  [see Figure \ref{fig:g2_rho2_two_imp} ($a_1$)]. The latter fact can also be observed in the two-body density $\rho_{AB}^{(2)}$ for two particles of different species [see Figure \ref{fig:g2_rho2_two_imp} ($c_1$) ,($f_1$)].Indeed, to a large extent the two species avoid each other, which indicates a phase separation where the impurity species occupies two adjacent outer wells and the majority species populates the other unoccupied wells. Nevertheless, this phase-separation is not complete and there is a finite, but small, overlap of the two species which we already discussed in the framework of the von Neumann entropy [see Figure \ref{fig:gs_densities_two_imp} (a)].

Coming back to the question of how the individual impurities distribute, in Figure \ref{fig:g2_rho2_two_imp} ($b_1$) it becomes clear that the impurity species occupying two adjacent sites may either have two particles at the same site or on two different sites. In this sense, the impurities are delocalized over these two sites. We also note that we have a slightly increased probability of finding the impurities at the same site which is adjacent to the sites occupied by the majority species. 

In order to understand whether the findings in the two-body densities are dominated by two-body processes we subsequently investigate the noise-correlation for two particles of the same species (see equation \ref{eq:noise_correlation}) for $V_0/E_R=5.7$, $V_0/E_R=8.9$ and $g_{AB}/E_R\lambda=0.12$. For $V_0/E_R=8.9$ the structure of the noise-correlation for two particles of the A species [Figure \ref{fig:g2_rho2_two_imp} ($d_2$)] is similar to the one for a single impurity in Figure \ref{fig:g2_rho2_single_imp} ($a_2$) which is to be expected since the particles of the majority species are in a Mott insulator-like state. Interestingly, in this scenario the noise-correlation for two impurities only shows a very weak structure of the order of $10^{-2}$ \footnote{These are due to a very small population of the impurity species in the site adjacent to the opposite outer well.} [see Figure \ref{fig:g2_rho2_two_imp} ($e_2$)]. This means that the accumulation of the two impurities in a single site is well described by the one-body densities such that the measurement of one impurity is independent of the previous one of the other impurity. However, this is not the case for $V_0/E_R=5.7$ and $g_{AB}/E_R\lambda=0.12$, where the impurities accumulate in adjacent sites. Here, the conditional probability of finding two particles of the impurity species at the same position deviates strongly from the unconditional one [see Figure \ref{fig:g2_rho2_two_imp} ($b_2$)]. This is also the case for finding the impurities at different sites. Also, for two particles of the majority species we observe that $g_{AA}^{(2)}\neq0$ in the relevant occupied sites [see Figure \ref{fig:g2_rho2_two_imp} ($a_2$)], with a structure similar to Figure \ref{fig:g2_rho2_two_imp} ($d_2$). Therefore, we can conclude that for $V_0/E_R=5.7$ not only the particles of the A species exhibit two-body correlation effects but also the two impurities as compared to the case of $V_0/E_R=8.9$.

In summary, we have found that for large lattice depths and strong interspecies interaction strengths two impurities accumulate in a single outer lattice well, a behavior that is similar to the case of a single impurity.  As a result the majority species occupies the other four lattice sites in a Mott insulator-like state and the composite system forms a doped insulator configuration. However, for smaller lattice depths the impurity species distributes over two adjacent lattice sites which in turn leads to a larger overlap of the two species. Also, the majority species now dominantly populates the three remaining unoccupied wells. This change in the distribution is accompanied by the presence of non-negligible correlations not only among the majority species atoms, as in the previous case, but also between the impurity atoms.
\section{Correlated Tunneling Dynamics}
\label{sec:Tunneling_Dynamics}
\begin{figure*}[t]
	\includegraphics[width=\textwidth]{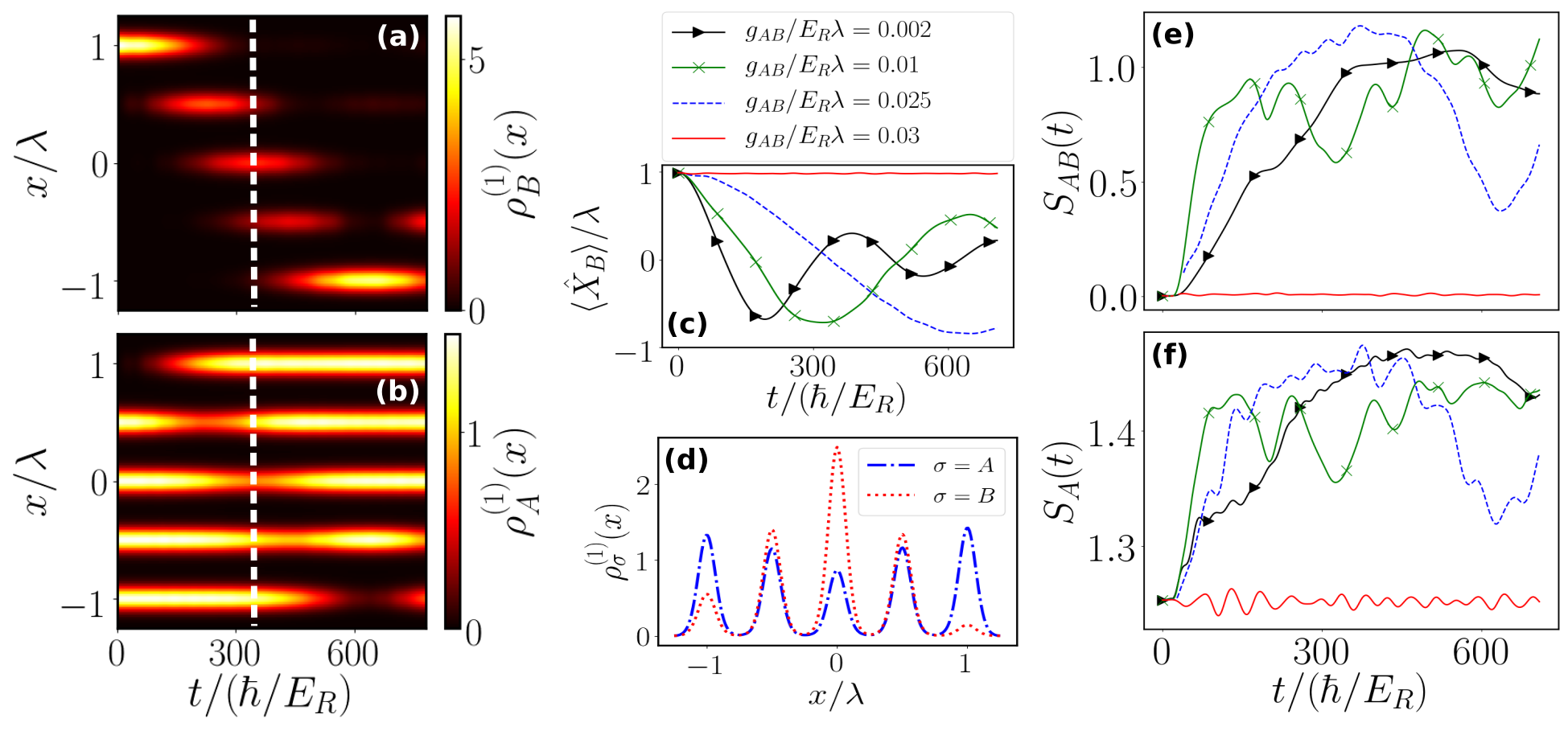}
	\caption{Temporal evolution of the one-body density (a) $\rho^{(1)}_B(x)$ of the species B and of (b) $\rho^{(1)}_A(x)$ of the species A upon quenching the interspecies interaction strength from $g_{AB}/E_R\lambda=0.04$ to $g_{AB}/E_R\lambda=0.025$. (c) Temporal evolution of the center of mass $\langle\hat{X}_B\rangle$ upon quenching to various $g_{AB}$ (see legend). (d) Density profiles of each species corresponding to the time indicated by the white line in panel (a) and (b), where the impurity is distributed the most in the lattice. Temporal evolution of (e) the von Neumann entropy $S_{AB}$ and (f) the fragmentation $S_A$ upon quenching to different $g_{AB}$ values (see legend). The particle number of the respective species is $N_A=4$ and $N_B=1$ and the lattice depth $V_0/E_R=10.5$.}
	\label{fig:trajectory_single_imp}
\end{figure*}
Having analyzed in detail the ground state properties of a lattice trapped bosonic mixture, we subsequently study its dynamical response upon quenching the interspecies interaction strength. To this end, we prepare our system, for one as well as two impurities, in its ground state for large lattice depths and strong interspecies interaction strengths such that the impurity species occupies one of the outer wells. In this regime the two species phase separate and form a particle-hole pair, as discussed in the previous sections \footnote{This selection of one of the two degenerate ground states can be performed by applying for example a small asymmetry to the lattice potential.}. By suddenly lowering the interspecies interaction strength we aim at initiating a tunneling process of the impurity species through the lattice geometry. We start by examining the tunneling properties in the case of a single impurity and then extend it to two impurities.  
\subsection{Transporting a Single Impurity}
\label{sec:Tunneling_Dynamics_Single_Impurity}
In the following we prepare our system in its ground state with a lattice depth $V_0/E_R=10.5$ and an interspecies interaction strength $g_{AB}/E_R\lambda=0.04$. This choice leads to a doped insulator density distribution of the two species as depicted in Figure \ref{fig:gs_densities_single_imp} (c), (d). In this sense the two species phase-separate and thereby form a particle-hole pair where the impurity plays the role of the particle. Moreover, the majority species atoms occupy each well separately. To trigger the dynamics, we quench the interspecies interaction strength to a smaller value such that we cross the transition from $S_{AB}\approx0$ to large values of $S_{AB}$ regarding the ground state entanglement crossover diagram shown in Figure \ref{fig:gs_densities_single_imp} (a). 

As a representative example of the emergent tunneling dynamics of each species in Figure \ref{fig:trajectory_single_imp} (a), (b) we present the temporal evolution of the corresponding one-body densities following a quench to $g_{AB}/E_R\lambda=0.025$, while keeping fixed $V_0/E_R=10.5$. In this case the impurity tunnels through the lattice geometry and ends up well localized in the opposite outer well [see Figure \ref{fig:trajectory_single_imp} (a)]. On the other hand the hole of the majority species in the initial well vanishes, which means that particles of the majority species travel towards this well which was initially solely populated by the impurity. Finally, for $t/(\hbar/E_R)\approx600$ a hole can again be found at the opposite outer well where now the impurity resides [see Figure \ref{fig:trajectory_single_imp} (b)].

\begin{figure*}[t]
	\includegraphics[width=\textwidth]{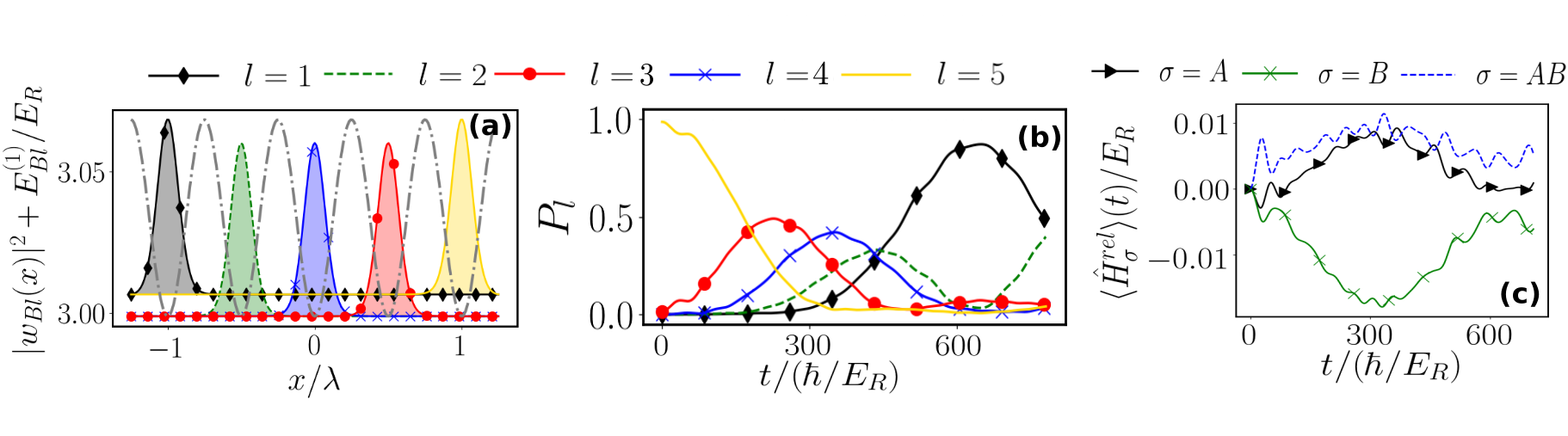}
	\caption{(a) Profile of the Wannier states associated with the lattice geometry. The offset of the probability density is given by the energy expectation value of $\hat{H}_{B}$ with respect to the respective Wannier states. The grey dash-dotted line represents the lattice. (b) Temporal evolution of the probability $P_l$ of finding the impurity in the $l-$th Wannier state upon quenching to $g_{AB}/E_R\lambda=0.025$. (c) Temporal evolution of the relative energy of the subsystems $\sigma$ as defined in equation \ref{eq:energy}. The particle number of the respective species is $N_A=4$ and $N_B=1$ and the lattice depth $V_0/E_R=10.5$.}
	\label{fig:single_wannier_single_imp}
\end{figure*}

In order to appreciate the involved tunneling processes we next examine the probability of finding an impurity particle in a specific Wannier state. To this end, we construct the operator
\begin{equation}
\hat{O}^{(1)}_l=|w^{B}_l\rangle\langle w^{B}_l|,
\label{eq:w_proj}
\end{equation}
where $|w^{B}_l\rangle\langle w^{B}_l|$ projects the particle of the B species onto the $l-$th Wannier state of the lowest band. Evaluating this operator with respect to the complete many-body wave function yields the probability $P_l=\langle\Psi_{\text{MB}}|\hat{O}^{(1)}_l|\Psi_{\text{MB}}\rangle$ of detecting an impurity particle in the $l-$th Wannier state.  Note that the complete many-body wave function is obtained via ML-MCTDHX and subsequently we analyze this high-dimensional object by evaluating $\hat{O}^{(1)}_l$ with respect to the wave function. In the following the Wannier states are ordered from left to right, i.e. $|w_5\rangle$ describes the Wannier state which is associated with the initially ($t=0$) populated well. The Wannier states prove to be a suitable basis set, since in all cases analyzed in the following, we find that $\sum_l P_l\approx99.9\%$.
Figure \ref{fig:single_wannier_single_imp} (b), shows the probability $P_l$ of finding the impurity in the $l-$th Wannier state upon quenching to $g_{AB}/E_R\lambda=0.025$. The energetically ordered associated Wannier states are depicted in Figure \ref{fig:single_wannier_single_imp} (a). The probability of finding the impurity in the initially occupied well decreases in the course of time to zero, while the other wells are populated such that at the end of the process a maximum probability in the left Wannier state occurs. This behavior clearly leaves an imprint on the relative energy of the subsystems $\sigma$
\begin{equation}
\langle\hat{H}^{rel}_\sigma\rangle(t)=\langle\hat{H}_\sigma\rangle(t)-\langle\hat{H}_\sigma\rangle(t=0),
\label{eq:energy}
\end{equation}
which we define as the deviation of the expectation values of the individual Hamiltonians $\hat{H}_\sigma$ (with $\sigma \in \{A,B,AB\}$) deviating from their initial value at $t=0$. 
The energy of the impurity first decreases during the transport to the opposite outer well and then increases again [see Figure \ref{fig:single_wannier_single_imp} (c)]. This can be easily understood in terms of the population of the corresponding Wannier states. The energy of the three central Wannier states is smaller as compared to the two outer ones [see Figure \ref{fig:single_wannier_single_imp} (a)]. Since the former are strongly populated in a superposition in the course of time, we observe a decrease of the impurity energy. The revival of the energy is in turn associated with the population of the other outer well which has again an increased energy (due to the hard-wall boundary conditions). The reduction of the impurity energy is accompanied by an increase of the energy of the majority species and the interspecies energy. The reason for the latter will be discussed below.

First, let us turn back to the superposition of the three central Wannier states in the course of time.
Interestingly, the propagation of the impurity species cannot be understood as a subsequent tunneling from one well to the adjacent one. The impurity rather delocalizes over the lattice geometry with a maximum delocalization indicated by the white dashed line in Figure \ref{fig:trajectory_single_imp} (a), (b). The one-body density for this scenario is depicted in Figure \ref{fig:trajectory_single_imp} (d). Here, the impurity strongly localizes in the central well, with around half of that population in the two directly adjacent sites and a minor density in the outer wells. Apparently, in order to avoid overlap, the majority species exhibits a density minimum in the central well with increasing density towards the outer wells. This can also be observed in the population of the corresponding Wannier states [see Figure \ref{fig:single_wannier_single_imp} (b)]. It is also worth noticing, that the one-body densities of both species clearly deviate from the ones discussed in the context of the ground states in Figure \ref{fig:gs_densities_single_imp} (b) where the roles of the two species are reversed with respect to the distribution of the one-body densities. 
However, this increased overlap not only increases the interspecies energy [see Figure \ref{fig:single_wannier_single_imp} (c)] but additionally leads to a drastic increase of the von Neumann entropy as in the static case for the ground states [see Figure \ref{fig:trajectory_single_imp} (e) dashed blue line]. Initially, we start with a von Neumann entropy of $S_{AB}\approx0$ which then increases to a maximum and decreases again when the impurity species resides in the outermost well. This decrease is due to the fact that the two species again phase-separate at later evolution times. Let us remark here that $S_{AB}$ does not drop back to zero since a minor residual density of the impurity remains in the second well. 

This tunneling process also has an impact on the correlations among the particles of the majority species. To unravel the role of these correlations, we define a measure for the correlations which are present in each subsystem itself. The spectral decomposition of the one-body density of species $\sigma$ reads
\begin{equation}
\rho_\sigma^{(1)}(x,t) = \sum_j n_{\sigma j}(t) \Phi^{*}_{\sigma j}(x,t)\Phi_{\sigma j}(x,t),
\label{eq:natural_populations}
\end{equation}
where $n_{\sigma j}(t)$ in decreasing order, obeying $ \sum_j n_{\sigma j}=1$, are the so-called natural populations and $\Phi_{\sigma j}(x,t)$ the corresponding natural orbitals. In this sense, the natural orbitals are the eigenstates, while the natural populations are the corresponding eigenvalues \cite{mlb1,mlb2,meyer_improved}, which are determined by diagonalizing the one-body density matrix. Similar to the Schmidt coefficients the natural populations serve as a measure for the correlations in a subsystem. In this spirit, we define the fragmentation \cite{lode1,lode2,Keiler_tunel} in the subsystem $\sigma$ as
\begin{equation}
S_{\sigma}(t)=-\sum_{j} n_{\sigma j}(t) \ln(n_{\sigma j}(t)).
\label{eq:fragmentation}
\end{equation}
Here, the case of $S_{\sigma}=0$ means that the subsystem $\sigma$ is not depleted, implying that all particles occupy the same single particle state, i.e. $n_{\sigma 1}=1$.
As just discussed, the delocalization of the impurity species is accompanied by a delocalization of the majority species. We observe that this is also reflected in the fragmentation $S_A$ of the majority species [see Figure \ref{fig:trajectory_single_imp} (f)]. At $t=0$ the majority species already exhibits a non-negligible fragmentation which is due to the fact that this species resides in a Mott insulator-like state. During the transport process, the majority species delocalizes which leads to an increase of $S_A$, thereby maximizing the latter. The subsequent decrease of the fragmentation can be associated with the transition back to a Mott insulator-like state. Since this transition is not complete due to residual densities in the wells, the fragmentation does not drop back to its initial value.

In order to clarify that this tunneling process is not unique to a specific post-quench interspecies interaction strength, we show in Figure \ref{fig:trajectory_single_imp} (c) the temporal evolution of the center of mass of the impurity which is defined as
\begin{equation}
\langle\hat{X}_B\rangle=\int_{-L/2}^{L/2}dx\rho_{B}^{(1)}(x) x.
\end{equation}
The initial value of the impurity's center of mass is given by $\langle\hat{X}_B\rangle/\lambda\approx1$, while the final position when the impurity is completely transported to the opposite outer well is $\langle\hat{X}_B\rangle/\lambda\approx-1$. For various post-quench $g_{AB}$ we find that the impurity can be transported to the opposite outer well, while the time it takes for the impurity to reach that well decreases with smaller $g_{AB}$. We also note that the impurity tunnels back to the initially occupied site afterwards, which shall not be the focus of further discussions. 
Nevertheless, it becomes additionally clear that it needs a certain minimal post-quench interspecies interaction strength in order to observe the impurity tunneling within the depicted time interval. E.g. for post-quench $g_{AB}/E_R\lambda=0.03$ the impurity rather resides in the initially occupied well, instead of tunneling through the lattice geometry, whereas for smaller $g_{AB}$ the tunneling process takes place.
As in the case of $g_{AB}/E_R\lambda=0.025$, which we discussed in detail, also for smaller $g_{AB}$ the von Neumann entropy as well as the fragmentation of the majority species are affected by the tunneling of the two species. However, the well separated initial increase and the subsequent decrease in both measures become less clear or not even evident anymore, which indicates that also the correlated tunneling takes place in a less structured manner with decreasing post-quench $g_{AB}$.

\begin{figure*}[t]
	\includegraphics[width=0.7\textwidth]{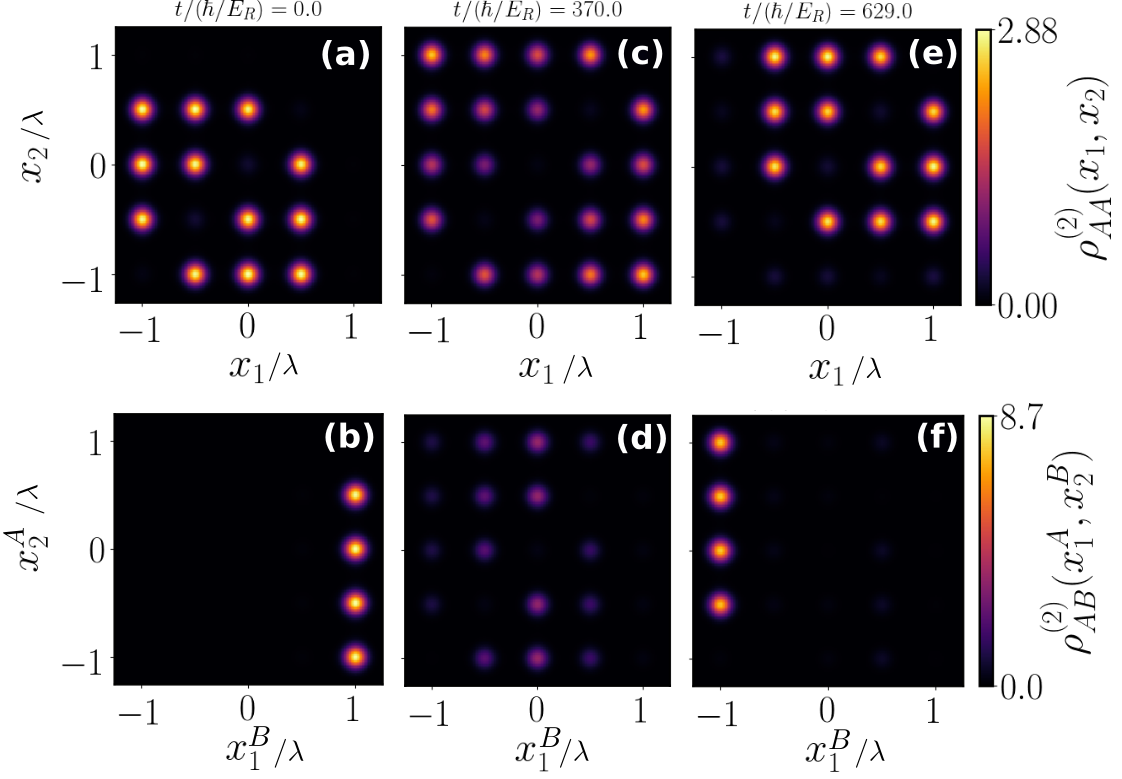}
	\caption{Snapshots of the two-body density for (a)-(c) two particles of the majority species $\rho^{(2)}_{AA}$ and (d)-(e) for one particle of the majority species and one impurity $\rho^{(2)}_{AB}$ upon quenching to $g_{AB}/E_R\lambda=0.025$. The first column corresponds to $t=0$, the second to $t/(\hbar/E_R)=370$ and the third to $t/(\hbar/E_R)=629$. The particle number of the respective species is $N_A=4$ and $N_B=1$ and the lattice depth $V_0/E_R=10.5$.}
	\label{fig:rho2_dynamics_single_imp}
\end{figure*}

In order to understand the above-described difference with varying post-quench $g_{AB}$, we first analyze the temporal evolution of the two-body densities for an interspecies interaction strength to $g_{AB}/E_R\lambda=0.025$. These observables provide a more comprehensive insight into the involved correlated processes indicated by the von Neumann entropy and the fragmentation of the majority species. 
In Figure \ref{fig:rho2_dynamics_single_imp} we present temporal snapshots of $\rho_{AB}^{(2)}$ and $\rho_{AA}^{(2)}$. The first column corresponds to $t=0$, while the second one refers to the case where the impurity species strongly delocalizes over the lattice geometry, thereby maximizing $S_{AB}$. The third column represents the time instance when the impurity resides in the opposite outer well with $\langle\hat{X}_B\rangle/\lambda\approx-1$. 
Focusing now on the majority species we clearly identify that throughout the tunneling process two particles of this species avoid to occupy the same site. Even in the case of maximum delocalization of the majority species two particles do not reside in the same site. We can understand this process as a superposition of all number states where four particles are distributed separately among the five Wannier states, e.g. $|1,1,1,1,0\rangle_A$, $|1,1,1,0,1\rangle_A$ etc. . In this sense, the hole in the majority species is delocalized over the lattice geometry. 
Moreover, we can infer that measuring the impurity at a specific site we will most probably not find a majority species atom at the same site. Even in the case of maximum delocalization these two particles will avoid each other. In this manner, the increase of $S_{AB}$ in the context of the delocalization is not a trivial single-particle effect. It is rather due to the strong avoidance of the two species and thereby the involved superposition of many contributing states, e.g. $|1,1,1,1,0\rangle_A\otimes|0,0,0,0,1\rangle_B$, $|1,1,1,0,1\rangle_A\otimes|0,0,0,1,0\rangle_B$ etc., which leads to the increase of the von Neumann entropy. This in turn leads to the delocalization of the two species on the level of the one-body densities when integrating out the respective degrees of freedom (see equation \ref{eq:obd}). 
Based on these findings we have strong indications that the tunneling of the two species manifests itself in the effective transport of a particle hole-pair from one of the outer wells to the opposite one. 

\begin{figure*}[t]
	\includegraphics[width=0.7\textwidth]{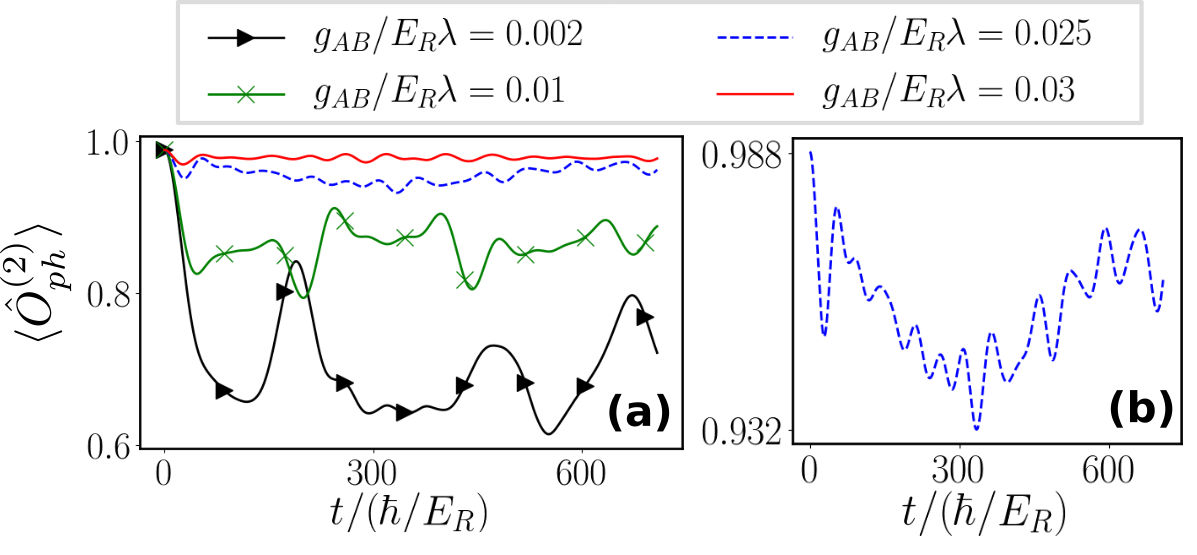}
	\caption{(a) Temporal evolution of the probability $\langle\hat{O}^{(2)}_{ph}\rangle$ of finding a particle hole pair in any lattice site for various post-quench interspecies interaction strengths (see legend). (b) Magnification of $\langle\hat{O}^{(2)}_{ph}\rangle$ for $g_{AB}/E_R\lambda=0.025$. The particle number of the respective species is $N_A=4$ and $N_B=1$, while the lattice depth is $V_0/E_R=10.5$.}
	\label{fig:particle_hole_single_imp}
\end{figure*}

To further quantify this tunneling  process we determine the probability of finding a particle hole pair in any lattice site in the course of time. We define this as the expectation value of the operator
\begin{equation}
\hat{O}^{(2)}_{ph}=\mathds{1}_A\otimes\mathds{1}_B-\frac{1}{N_AN_B}\sum_{l=1}^{5} \sum^{N_A N_B}_{ij}|w^{jB}_l\rangle\langle w^{jB}_l| \otimes  |w^{iA}_l\rangle\langle w^{iA}_l|,
\label{eq:particle_hole}
\end{equation}
with respect to the system's wave function, i.e. $\langle\hat{O}^{(2)}_{ph}\rangle$.
Here, $|w^{jB}_l\rangle\langle w^{jB}_l|$ projects the $j-$th particle of the B species onto the $l-$th Wannier state and $|w^{iA}_l\rangle\langle w^{iA}_l|$ projects the $i-$th particle of the A species onto the $l-$th Wannier state.
$\mathds{1}_\sigma$ are the unity operators of the respective subsystems $\sigma$.
Figure \ref{fig:particle_hole_single_imp} (a) presents the temporal evolution of the probability of finding a particle hole pair in any lattice site for various post-quench interspecies interaction strengths. 
Since our system is prepared in its ground state for all post-quench $g_{AB}$, initial particle-hole pair probability of $\langle\hat{O}^{(2)}_{ph}\rangle(t=0)\approx98.8\%$ holds. This means that we can safely assume that our system exhibits a particle-hole pair.

Focusing on the case of $g_{AB}/E_R\lambda=0.025$, which corresponds to the two-body densities depicted in Figure \ref{fig:rho2_dynamics_single_imp}, the quench leads to a slight reduction of $\langle\hat{O}^{(2)}_{ph}\rangle$ which recovers again for larger times. Closely inspecting the particle hole-pair probability [see Figure \ref{fig:particle_hole_single_imp} (b)], we can deduce that the strongest decrease happens at the time interval in which the two species strongly delocalize over the lattice geometry. For this reason, we can assume that our system is not solely described by the superposition of number states exhibiting a particle-hole pair, e.g. $|1,1,1,1,0\rangle_A\otimes|0,0,0,0,1\rangle_B$, $|1,1,1,0,1\rangle_A\otimes|0,0,0,1,0\rangle_B$ etc. . Nevertheless, the admixture of other states is rather small since in the case with the largest deviations we still find a particle hole-pair with a probability of $\langle\hat{O}^{(2)}_{ph}\rangle\approx93.2\%$. Moreover, we observe that $\langle\hat{O}^{(2)}_{ph}\rangle$ increases again in the course of time, if not completely to its initial value, which is associated with the impurity then occupying dominantly the opposite outer well. 
In contrast, when quenching to smaller values of $g_{AB}$, the stability of the particle-hole pair cannot be guaranteed anymore. For a quench to $g_{AB}/E_R\lambda=0.01$ we lose up to $~20\%$ of the particle-hole pair which becomes even more for a quench close to zero, i.e $g_{AB}/E_R\lambda=0.002$. In the latter case $\langle\hat{O}^{(2)}_{ph}\rangle$ drastically decreases to $~65\%$ in the first few time steps. In this sense, in order to maintain a stable particle-hole pair during the transport of the impurity to the opposite outer well it is necessary not to quench too strongly. On the other hand a quench which is too weak allows for a stable particle-hole pair, which resides in the well initially occupied by the impurity, e.g. $g_{AB}/E_R\lambda=0.03$ [see Figure \ref{fig:particle_hole_single_imp} (a) red line].

In conclusion, we have found that quenching the binary mixture starting in a phase separated state leads to a very controlled correlated tunneling dynamics. On the one hand the majority species tunnels such that its particles do not occupy the same lattice site. On the other hand we observe the effective transport of a particle-hole pair from one of the outer wells to the opposite one whose stability strongly depends on the presence of a finite interspecies interaction strength. Moreover, the tunneling dynamics is accompanied by a strong entanglement between the species as well as correlations among the majority species atoms.

\subsection{Transport Properties of Two Bosonic Impurities}
\label{sec:Tunneling_Dynamics_Two_Impurities}
\begin{figure*}[t]
	\includegraphics[width=0.9\textwidth]{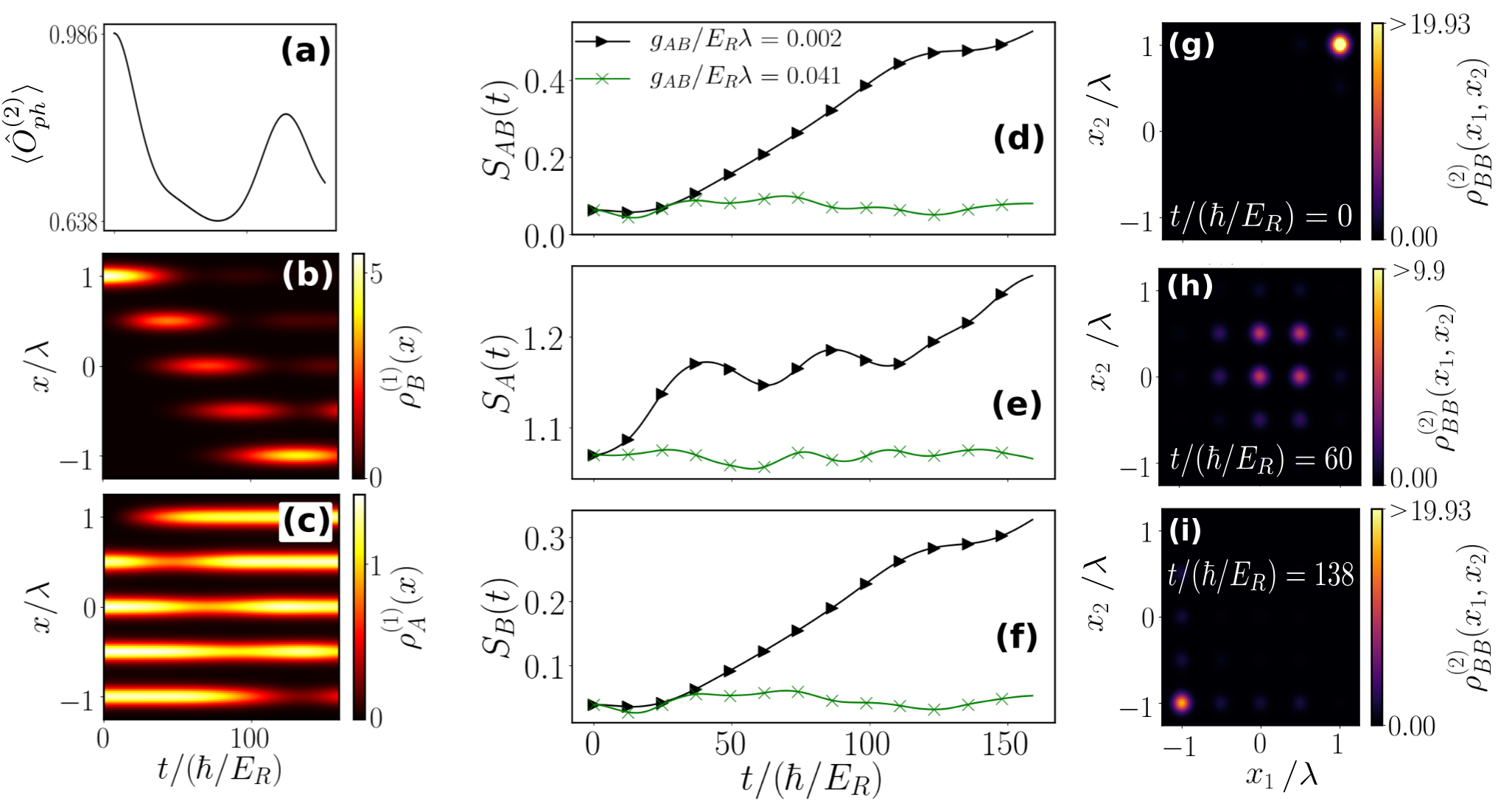}
	\caption{(a) Temporal evolution of the probability $\langle\hat{O}^{(2)}_{ph}\rangle$ of finding a particle hole-pair in any lattice site for a post-quench $g_{AB}/E_R\lambda=0.002$. Temporal evolution of the one-body density (b) $\rho^{(1)}_B(x)$ of the species B and of (c) $\rho^{(1)}_A(x)$ of the species A upon quenching the interspecies interaction strength from $g_{AB}/E_R\lambda=0.05$ to $g_{AB}/E_R\lambda=0.002$. Temporal evolution of (d) the von Neumann entropy $S_{AB}$, (e) the fragmentation $S_A$ and (f) the fragmentation $S_B$ upon quenching to various $g_{AB}$. (g)-(i) Snapshots of the two-body density of the impurities $\rho^{(2)}_{BB}$ upon quenching to $g_{AB}/E_R\lambda=0.002$. The particle number of the respective species is $N_A=4$ and $N_B=2$ and we set the lattice depth to $V_0/E_R=8.9$.}
	\label{fig:trajectory_two_imp}
\end{figure*}
Having understood the correlated tunneling dynamics of a single impurity interacting repulsively with a majority species we now turn to the case of two non-interacting bosonic impurities. We prepare our system in its ground state characterized by a lattice depth $V_0/E_R=8.9$ and interspecies interaction strength $g_{AB}/E_R\lambda=0.05$, while $g_{AA}/E_R\lambda=0.04$. This leads to a density distribution of the two species as depicted in Figure \ref{fig:gs_densities_two_imp} (e), (f), such that the two species phase-separate. Moreover, the two impurities accumulate at the same single lattice site, while the majority species atoms strongly avoid each other, occupying the lattice sites each one separately. In this sense, our initial state is similar to the one with a single impurity, apart from the additionally added impurity. 

Quenching the interspecies interaction strength to $g_{AB}/E_R\lambda=0.002$ we observe a tunneling of the impurity species to the opposite outer well [see Figure \ref{fig:trajectory_two_imp} (b)]. However, compared to the case of a single impurity a larger portion of density remains in the wells which are traversed during the dynamics. This means that compared to the single impurity case the transport portion of two impurities is less complete due to some density remaining in the traversed lattice sites. In turn the majority species tunnels in a counterflow into the opposite direction as compared to the direction of the impurities. In this context, one might again ask whether this process can be described by an effective transport of a particle hole-pair. Figure \ref{fig:trajectory_two_imp} (a) shows the corresponding particle hole-pair probability $\langle\hat{O}^{(2)}_{ph}\rangle$ during the evolution. Initially, the particle hole-pair process is the dominant tunneling channel with a probability of $\langle\hat{O}^{(2)}_{ph}\rangle\approx98.6\%$ which later on drastically decreases to $\langle\hat{O}^{(2)}_{ph}\rangle\approx63.8\%$. The subsequent increase of this probability, which is associated with the impurities residing in the opposite outer well, does not revive to its initial value, indicating the loss of the particle hole-pair. This effect is similar to the one discussed for a single impurity when quenching to very small values of $g_{AB}$, where the stability of the particle hole-pair cannot be guaranteed either.
Apart from that, the system still exhibits a rather pronounced increase of correlations in the course of time. This involves for example the increase of the von Neumann entropy which can be traced back to the residual impurity density in the remaining wells during their transport to the other side of the lattice geometry. As a result the spatial overlap between the species is increased and thereby also $S_{AB}$ [see Figure \ref{fig:trajectory_two_imp} (d)]. Furthermore the initially uncorrelated impurity pair, $S_B(t=0)\approx0$, develops correlations during the propagation to the opposite outer well which can be attributed to the incomplete transfer of the impurities to this very site [see Figure \ref{fig:trajectory_two_imp} (e)]. Similarly, the increase of $S_A$ is attributed to the incomplete transfer of the effective hole into the opposite direction (as compared to the propagation direction of the impurities) [see Figure \ref{fig:trajectory_two_imp} (f)].

Furthermore, it is of interest to analyze in which manner the two bosonic impurities propagate in relation to each, i.e. whether they move as a pair or they delocalize in the course of time. In order to answer this question we inspect the two-body density of the impurities $\rho^{(2)}_{BB}$ upon quenching to $g_{AB}/E_R\lambda=0.002$ [see Figure \ref{fig:trajectory_two_imp} (g)-(i)]. The initially accumulated (in an outer well) impurities delocalize over next-neighbour sites, which means that it is possible for the two impurities to either reside at the same site or in adjacent ones [see Figure \ref{fig:trajectory_two_imp} (h)]. However, during the propagation there are time instances at which the probability of finding the impurities at the same site is more pronounced as compared to detecting them in adjacent ones (not shown here). From this we can conclude that in general the impurities delocalize during the propagation until they reach the opposite outer well where they eventually strongly localize [see Figure \ref{fig:trajectory_two_imp} (i)].

Let us finally remark that in comparison to the single impurity case, for two impurities it is necessary to quench to lower interspecies interaction strengths in order to achieve a significant transport of the impurity species to the opposite outer well. For smaller quench amplitudes we either find no tunneling of the two species [cf. Figure \ref{fig:trajectory_two_imp} (d)-(f) green crosses] or the impurities delocalize over the lattice geometry without accumulating in the opposite outer well. Therefore, a transport of the impurity species to the opposite outer well as in the case of a single impurity is only possible for large quench amplitudes. However, this leads to a loss of the initial particle-hole pair and a less structured development of correlations.

\section{Conclusions and Outlook}
\label{sec:Conclusion}
\begin{figure*}[t]
	\centering
	\includegraphics[width=0.8\textwidth]{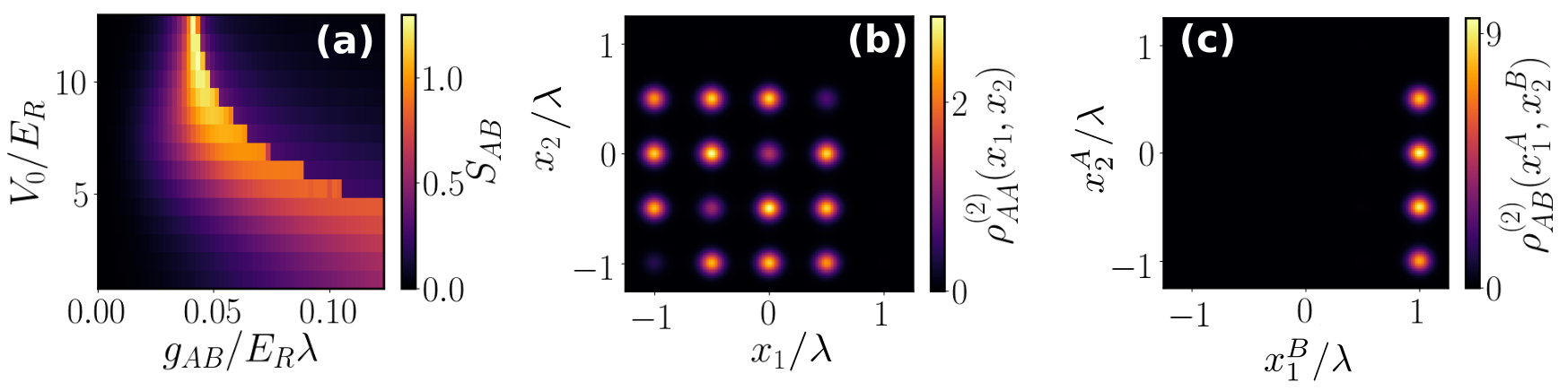}
	\caption{(a) The von Neumann entropy $S_{AB}$ as a function of the interspecies interaction strength $g_{AB}$ and the lattice depth $V_0$. The two-body density for (b) two particles of the majority species $\rho^{(2)}_{AA}$ and (c) for one particle of the majority species and one impurity $\rho^{(2)}_{AB}$.  We consider $V_0/E_R=10.5$ and $g_{AB}/E_R\lambda=0.12$. The particle number of the respective species is $N_A=5$ and $N_B=1$.}
	\label{fig:single_imp_incom}
\end{figure*}
We have investigated the ground state properties of a binary bosonic mixture trapped in a one-dimensional lattice geometry consisting of a majority species which is doped by one or two bosonic impurities. We demonstrate the existence of a crossover diagram of the interspecies entanglement as a function of the lattice depth and the interspecies interaction strength. The transition from strong interspecies entanglement to $S_{AB}=0$ is accompanied by a crossover from a spatially delocalized impurity species to its strong localization in one of the outer lattice wells. For large lattice depths we find this transition for the case of a single as well as two bosonic impurities. Moreover, analyzing the corresponding two-body densities we can conclude that the majority species occupies a Mott insulator-like state, while the two species phase-separate, thereby forming a particle hole-pair. For two impurities we additionally observe in the case of smaller lattice depths a localization of the impurity species in two adjacent sites. This phenomenon also manifests itself in the impurity-impurity correlations, which we do not find when the impurities localize in a single site.

Having understood the ground state distributions, in the next step we aimed at transporting the impurity species through the lattice by performing an interspecies interaction quench from a Mott to the superfluid phase of the composite system. For a single impurity the initial particle hole-pair tunnels to the opposite outer lattice well, while we are able to control the stability of the particle hole-pair by varying the post-quench interspecies interaction strength. Here, it is important to perform weak amplitude quenches in order to maintain the initial particle hole-pair.
However, for two impurities we cannot guarantee a stable transport of this pair. This is due to the fact that only large quench amplitudes can initiate a significant transport of the impurities to the opposite outer well at all. As in the case of a single impurity such large quench amplitudes lead to a strong loss of the particle hole-pair probability.

The understanding of the crossover between a spatially delocalized to a localized bosonic impurity species serves as a perfect starting point for even more complex setups, e.g. by introducing more lattice atoms and sites.
Indeed, there are several possible directions of future investigations. For example, it would be of interest to allow for a spin degree of freedom in both species, such that the particle hole-pair may carry an additional effective spin. In this spirit, a quench of the interspecies interaction strength might lead to a redistribution of the spins in the system or an effective spin transport. Another fruitful perspective is to investigate the ground state phase diagram of the setting considered herein but including dipolar interactions within and between the species. In this way, it would be possible to generate more phases due to the long-range character of the interactions. 

\appendix
\section{Doping a unit-filling bosonic majority species}
\label{sec:Appendix1}
To explicitly showcase the generalization of our findings regarding the ground state properties of the considered lattice trapped bosonic mixture, we consider a unit filling of the majority species, doping it with a single impurity, i.e. $N_A=5$ and $N_B=1$. We explore the ground state of the system  for varying $V_0$ and $g_{AB}$, while fixing the intraspecies interaction strength to $g_{AA}/E_R \lambda=0.04$.
Figure \ref{fig:single_imp_incom} (a) shows the von Neumann entropy of the ground state as a function of the interspecies interaction strength $g_{AB}$ and the lattice depth $V_0$. As in the case of $N_A=4$ and $N_B=1$ [Figure \ref{fig:gs_densities_single_imp} (a)], we find a region of increased interspecies entanglement, which decreases to zero for a further increase of $g_{AB}$. This is again associated with a strong delocalization of both species for large $S_{AB}$ and a phase separation in case of $S_{AB}\approx0$ occurring for large $g_{AB}$. However, there is a slight change compared to $N_B=4$ in the corresponding two-body densities for $V_0/E_R=10.5$ and $g_{AB}/E_R\lambda=0.12$, where an interspecies phase-separation takes place. Since the majority species exhibits an additional particle, the latter distributes over the four lattice sites which are occupied by the majority species in case of an interspecies phase-separation. For this reason a non-negligible, but small, probability of finding two particles of the A species at the same site occurs. 
In  $\rho^{(2)}_{AB}$ the phase separation is still reflected with an increased probability of finding the majority species in the two central occupied sites, which is due to the additional majority species particle.
In this sense, also for $N_A=5$ we find the formation of a particle hole-pair for large $g_{AB}$, while the additional particle in the majority species delocalizes over the sites occupied by the latter.

\acknowledgements
P. S. gratefully acknowledges funding by the Deutsche Forschungsgemeinschaft in the framework of the SFB 925 "Light induced dynamics and control of correlated quantum systems". K. K. gratefully acknowledges a scholarship of the Studienstiftung des deutschen Volkes. S. I. M gratefully acknowledges financial support in the framework of the Lenz-Ising Award of the University of Hamburg. \par

\end{document}